\documentclass[sigconf]{aamas} 

\usepackage{balance} 
\usepackage{amsmath,amsfonts}
\usepackage{graphicx}
\usepackage{textcomp}
\usepackage{xcolor}
\usepackage{comment}
\usepackage[hang,flushmargin]{footmisc}
\usepackage{lipsum}
\makeatletter
\newcommand{\algorithmfootnote}[2][\footnotesize]{%
  \let\old@algocf@finish\@algocf@finish
  \def\@algocf@finish{\old@algocf@finish
    \leavevmode\rlap{\begin{minipage}{\linewidth}
    #1#2
    \end{minipage}}%
  }%
}
\makeatother

\usepackage{algpseudocode,algorithm}
\usepackage{bm}
\usepackage{url}
\usepackage{subfigure}

\newcommand{\argmax}{\mathop{\rm arg~max}\limits}

\newcommand{\sign}{\mathop{\rm sign}\limits}
\usepackage{amsthm}
\theoremstyle{definition}

\newtheorem{problem}{Problem}

\newtheorem{definition}{Definition}

\usepackage{xparse}

\NewDocumentCommand{\timeseries}{O{}O{}O{}}{#1_{#2}[{#3}]}
\NewDocumentCommand{\ret}{O{}O{}}{\timeseries[r][#1][#2]}
\NewDocumentCommand{\vecret}{O{}O{}}{\bm{r}_{#1}[{#2}]}
\NewDocumentCommand{\predret}{O{}O{}}{\hat{r}_{#1}[{#2}]}
\NewDocumentCommand{\signret}{O{}O{}}{\hat{b}_{#1}[{#2}]}
\NewDocumentCommand{\price}{O{}O{}}{X_{#1}[{#2}]}
\NewDocumentCommand{\numterm}{}{M}
\usepackage{footnote}
\makesavenoteenv{tabular}

\setcopyright{ifaamas}
\acmConference[AAMAS '21]{Proc.\@ of the 20th International Conference on Autonomous Agents and Multiagent Systems (AAMAS 2021)}{May 3--7, 2021}{Online}{U.~Endriss, A.~Now\'{e}, F.~Dignum, A.~Lomuscio (eds.)}
\copyrightyear{2021}
\acmYear{2021}
\acmDOI{}
\acmPrice{}
\acmISBN{}

\acmSubmissionID{445}

\title[Trader-Company Method]{Trader-Company Method: A Metaheuristic for Interpretable Stock Price Prediction}

\author{Katsuya Ito}
\affiliation{
\institution{Preferred Networks, Inc.}
}
\email{katsuya1ito@preferred.jp}

\author{Kentaro Minami}
\affiliation{
\institution{Preferred Networks, Inc.}
}
\email{minami@preferred.jp}

\author{Kentaro Imajo}
\affiliation{
\institution{Preferred Networks, Inc.}
}
\email{imos@preferred.jp}

\author{Kei Nakagawa}
\affiliation{
\institution{Nomura Asset Management Co., Ltd.}
}
\email{k-nakagawa@nomura-am.co.jp}

\begin{abstract}
Investors try to predict returns of financial assets to make successful investment.
Many quantitative analysts have used machine learning-based methods to find unknown profitable market rules from large amounts of market data.
However, there are several challenges in financial markets hindering practical applications of machine learning-based models.
First, in financial markets, there is no single model that can consistently make accurate prediction because traders in markets quickly adapt to newly available information. Instead, there are a number of ephemeral and partially correct models called ``alpha factors''.
Second, since financial markets are highly uncertain, ensuring interpretability of prediction models is quite important to make reliable trading strategies.
To overcome these challenges, we propose the Trader-Company method, a novel evolutionary model that mimics the roles of a financial institute and traders belonging to it.
Our method predicts future stock returns by aggregating suggestions from multiple weak learners called Traders. A Trader holds a collection of simple mathematical formulae, each of which represents a candidate of an alpha factor and would be interpretable for real-world investors. The aggregation algorithm, called a Company, maintains multiple Traders. By randomly generating new Traders and retraining them, Companies can efficiently find financially meaningful formulae whilst avoiding overfitting to a transient state of the market.
We show the effectiveness of our method by conducting experiments on real market data.
\end{abstract}


\newcommand{\BibTeX}{\rm B\kern-.05em{\sc i\kern-.025em b}\kern-.08em\TeX}

\begin{document}


\pagestyle{fancy}
\fancyhead{}


\maketitle 

\section{Introduction}
Developing quantitative trading strategies is a universal task in the financial industry \cite{marcoslopezdeprado2018}.
Many quantitative models have been proposed to predict the behavior of financial markets \cite{McLean2012,Harvey2015,Wiecki2016}.
For example, Fama--Frech's three-factor model and five-factor model \cite{fama1992,fama2015,Carhart1997} have been standard asset pricing models for many years.
Technical indicators such as Moving Average Convergence Divergence (MACD) and Relative Strength Index (RSI) are also prediction methods that have been used by many traders \cite{stevenachelis2000,Li2019}.

Although many quantitative analysts are struggling to derive new rules from newly available big data, there has been no gold-standard practical method that can fully leverage these data \cite{Wiecki2016}.
We believe that there are the following two challenges that are hindering the development of quantitative models today.

\begin{figure*}[ht]
  \includegraphics[width=\textwidth]{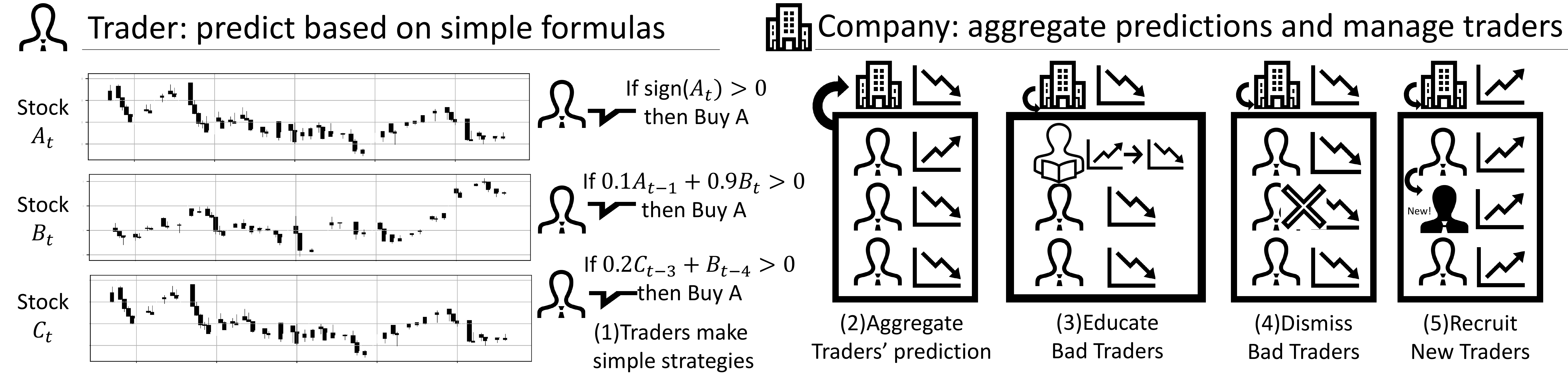}
  \caption{Illustration of our method.
  (1) Traders predict the return of assets using simple formulas. 
  Companies manage Traders and combine Traders' predictions into one value.
  The Company algorithm consists of four functions: (2) prediction by aggregation, (3) education of bad Traders, (4) dismissal of bad Traders, and (5) recruitment of new Traders.}
  \Description{Figure showing an illustration of the proposed method, the Trader--Company method. On the left-hand side, the figure shows three artificial examples of stock prices. On the middle, the figure shows the concept of Traders, which imitate human traders who make predictions by simple mathematical formulae. On the right-hand side, the figure shows the concept of Companies that have four functionalities: Aggregate Traders' prediction, Educate bad Traders, Dismiss bad Traders, and Recruit new Traders.}
  \label{paper_idea}
\end{figure*}

\subsection{Tackling Nearly-Efficient Markets}
\label{weekmodel}

Our first challenge is to tackle the non-stationary and noisy nature of the financial market, which is known as the \textit{efficiency} of markets. The widely acknowledged Efficient Market Hypothesis \cite{Fama1970} states that asset prices reflect all available information, correcting undervalued or overvalued prices into fair values. In an efficient market, investors cannot outperform the overall market because asset prices quickly follow other traders' strategic and adversarial activities \cite{LopezdePrado2015}. In fact, many empirical studies have reported that real-world markets are nearly efficient \cite{Cont2001,McLean2012,Wiecki2016}.
Due to this, future stock returns are hardly predictable in most markets, and no single explanatory model can consistently make an accurate prediction.

On the other hand, there still is a common belief that stock returns can be predictable in a sufficiently short time period, which suggests the existence of investments or trading strategies that beat the overall market at least temporarily.
In particular, potential sources of profitability would come from some simple mathematical formulae, called \textit{alpha factors} \cite{Kakushadze2018, tulchinsky2015alphas}. Typical alpha factors used in production are given as combinations of few elementary functions and arithmetic operations. For example,
\[
    \log (\text{yesterday's close price} / \text{yesterday's open price})
\]
represents the classical momentum strategy \cite{jegadeesh1993moment}. It has been reported that there is a variety of mathematical formulae with reasonably low mutual correlations \cite{Kakushadze2018}, each of which can be a good trading signal and actually usable in real-life trading.

Although the efficacy of a single formula is slight and ephemeral, combining multiple formulae in a sophisticated way can lead to a more robust trading signal. We hypothesize that we can overcome the instability and the uncertainty of markets by maintaining multiple ``weak models'' given as simple mathematical formulae. This is in the same spirit as the ensemble methods (see e.g., \cite{Timmermann2006, hastie2009elements}), but paying more attention to specific structure of real-world alpha factors may improve the performance of the resulting trading strategy.

\subsection{Interpretability of Trading Strategies}\label{interpret}

The second challenge is to gain the interpretability of models.
As mentioned above, it is hard to achieve consistently high performance in the financial markets.
Even if we could have the best possible trading strategies at hand, their predictive accuracies are quite limited and unsustainable.
To gain intuition, suppose that we forecast the rise or fall of a single stock. Then, the accuracy is typically no higher than 51\%, which is approximately the chance rate.
As such, investors should worry about their trading strategies having a large uncertainty in the returns and the risks.

In such a highly uncertain environment, the interpretability of models is of utmost importance.
Warren Buffett said, ``Risk comes from not knowing what you are doing'' \cite{Hagstrom1997}.
As his word implies, investors may desire the model to be explainable to understand what they are doing.
In fact, historically, investors and researchers have preferred linear factor models to explain asset prices (e.g., \cite{fama1992,fama2015}), which are often believed as interpretable.
On the other hand, for machine learning-based strategies, the lack of interpretability can be an obstacle to practical use, without which investors cannot understand their own investments nor ensure accountability to customers. We believe that using the combination of simple formulae mentioned in the previous subsection can open up a possibility of interpretable machine learning-based trading strategies.

\subsection{Our Contributions}
To address the aforementioned challenges, we propose the Trader-Company method, a new metaheuristics-based method for stock price prediction.
Our method is inspired by the role of financial institutions in the real-world stock markets.
Figure \ref{paper_idea} depicts the entire framework of our method. Our method consists of two main ingredients, \textit{Traders} and \textit{Companies}.
A single Trader predicts the returns based on simple mathematical formulae, which are postulated to be good candidates for interpretable alpha factors. 
Brought together by a Company, Traders act as weak learners that provide partial information helping the Company's eventual prediction. The Company also updates the collection of Traders by generating (i.e., hiring) new candidates of good Traders as well as by deleting (i.e., dismissing) poorly performing Traders. This framework allows us to effectively search over the space of mathematical formulae having categorical parameters.

We demonstrate the effectiveness of our method by experiments on real market data. We show that our method outperforms several standard baseline methods in realistic settings. Moreover, we show that our method can find formulae that are profitable by themselves and simple enough to be interpreted by real-world investors.

\section{Problem Definition}\label{sec:probdef}

In this section, we provide several mathematical definitions of financial concepts and formulate our problem setting.
Table \ref{tab:notation} summarizes the notation we use in this paper.
\subsection{Problem Setting}\label{sec:probset}
Our problem is to forecast future returns of stocks based on their historical observations. To be precise, let $\price[i][t]$ be the price of stock $i$ at time $t$, where $1 \le i \le S$ is the index of given stocks and $0 \le t \le T$ is the time index.
Throughout this paper, we consider the logarithmic returns of stock prices as input features of models. 
That is, we denote the one period ahead return of stock $i$ by \begin{equation}\label{logret}
\ret[i][t] := \log (\price[i][t] / \price[i][t-1]) \approx \frac{\price[i][t]-\price[i][t-1]}{\price[i][t-1]}.
\end{equation}
We denote returns over multiple periods and returns over multiple periods and multiple stocks by 
\begin{equation}\label{vecret}
\vecret[i][u:v]=(\ret[i][u],\cdots,\ret[i][v]),
\vecret[i:j][u:v]=(\ret[i][u:v],\cdots,\ret[j][u:v])
\end{equation}

Our main problem is formulated as follows.
\begin{problem}[one-period-ahead prediction]\label{main_problem}
The predictor sequentially observes the returns $\ret[i][t]$ ($1 \le i \le S$) at every time $0 \le t \le T$. For each time $t$, the predictor predicts the one-period-ahead return $\ret[i][t+1]$ based on the past $t$ returns $\vecret[1:S][0:t]$. That is, the predictor's output can be written as
\begin{equation}\label{predret}
\predret[i][t+1] = f_{t}(\vecret[1:S][0:t])
\end{equation}
for some function $f_t$ that does not depend on the values of $\vecret[1:S][t:T]$.
\end{problem}

\subsection{Evaluation}
To evaluate the goodness of the prediction, we use the cumulative return defined as follows. Given a predictor's output $\predret[i][t+1]$, we define the ``canonical'' trading strategy as
\begin{equation}
    \signret[i][t+1]
    = \sign (\predret[i][t+1]).
    \label{eq:canonical_strategy}
\end{equation}
Here, the value of $\signret[i][t]$ represents the trade of stock $i$ at time $t$. That is, $\signret[i][t] = \pm 1$ respectively means that the strategy buys/sells one stock and sell/buy it back after one periods.
Thus, equation \eqref{eq:canonical_strategy} can be interpreted as a strategy that buys a unit amount of a stock if its one period ahead return is expected to be positive and otherwise sells it. Then, we define the cumulative return of the prediction $\predret[i][t+1]$ as \[
\timeseries[C][i][t]
= \sum_{u = 0}^t \signret[i][u+1]\ret[i][u+1]
= \sum_{u = 0}^t \sign(\predret[i][u+1])\ret[i][u+1].
\]
If the predictor could perfectly predict the sign of the one period ahead returns, the above canonical strategy yields the maximum cumulative return among all possible strategies that can only trade unit amounts of stocks. As such, we consider $\timeseries[C][i][t]$ as an evaluation major of the prediction.

\begin{table}
\caption{Notation.}
\label{tab:notation}
\begin{center}

\begin{tabular}{clc}
\hline
\textbf{Notation} & \multicolumn{1}{c}{\textbf{Meaning}} & \textbf{Def.} \\
\hline
$\price[i][t]$ & 
\begin{tabular}{l}
stock price of stock $i$ at time $t$\\
where $1\le i \le S, 0\le t \le T$
\end{tabular}& $\S$ \ref{sec:probset} \\
$\ret[i][t]$ & logarithmic return of $i$ at $t$& \eqref{logret}\\
$\vecret[i][u:v]$ & $(\ret[i][u],\cdots,\ret[i][v])$ & \eqref{vecret}  \\
$\vecret[i:j][u:v]$ & $(\ret[i][u:v],\cdots,\ret[j][u:v])$ & \eqref{vecret}  \\
$\predret[i][t],\signret[i][t]$ & predicted value of $\ret[i][t]$ and its sign & \eqref{predret}\eqref{eq:canonical_strategy}\\
\begin{tabular}{l}$\numterm,P_j,Q_j,D_j$\\$F_j,A_j,O_j$\end{tabular}
& hyper-parameters of Traders & \eqref{traders}\\
\hline
\end{tabular}
\end{center}

\end{table}

\section{Trader-Company Method}

In this section, we present the \textit{Trader-Company method}, a new metaheuristics-based prediction algorithm for stock prices.

Figure \ref{paper_idea} outlines our proposed method.
Our method consists of two main components, \textit{Traders} and \textit{Companies}, which are inspired by the role of human traders and financial institutes, respectively. A Trader predicts the returns using a simple model expressing realistic trading strategies. A Company combines suggestions from multiple Traders into a single prediction. To train the parameters of the proposed system, we employ an evolutionary algorithm that mimics the role of financial institutes as employers of traders. During training, a Company generates promising new candidates of Traders and deletes poorly performing ones.
Below, we provide more detailed definitions and training algorithms for Traders and Company.

\subsection{Traders - Simple Prediction Module}
First, we introduce the Traders, which are the minimal components in our proposed framework.
As mentioned in the introduction, trading strategies used in real-life trading are made of simple formulae involving a small number of arithmetic operations on the return values \cite{Kakushadze2018}.
We postulate that there are a number of unexplored profitable market rules that can be represented by simple formulae, which leads us to the following definition of a parametrized family of formulae.

\begin{definition}
A Trader is a predictor of one period ahead returns defined as follows.
Let $\numterm$ be the number of terms in the prediction formula.
For each $1 \le j \le \numterm$, we define $P_j,Q_j$ as the indices of the stock to use, $D_j,F_j$ as the delay parameters, $O_j$ as the binary operator, $A_j$ as the activation function, and $w_j$ as the weight of the $j$-th term. 
Then, the Trader predicts the return value $\ret[i][t+1]$ at time $t + 1$ by the formula
\begin{equation}\label{traders}
    f_\Theta(\vecret[1:S][0:t])= \sum_{j=1}^\numterm w_j                 A_j(O_j(\ret[P_j][t-D_j],\ret[Q_j][t-F_j])).
\end{equation}
where $\Theta$ is the parameters of the Trader:
\[
    \Theta := \{\numterm,\{P_j,Q_j,D_j,F_j,O_j,A_j,w_j\}_{j=1}^\numterm\}.
\]
\end{definition}

For activation functions $A_j$, we use standard activation functions used in deep learning such as the identity function, hyperbolic tangent function, hyperbolic sine function, and Rectified Linear Unit (ReLU). 
For the binary operators $O_j$, we use several arithmetic binary operators (e.g., $x + y$, $x - y$, and $x \times y$), the coordinate projection, $(x, y) \mapsto x$, the max/min functions, and the comparison function $(x > y) = \sign(x - y)$.

Our definition of the Trader has several advantages.
First, the formula \eqref{traders} is ready to be interpreted in the sense that it has a similar form to typical human-generated trading strategies \cite{Kakushadze2018}.
Second, the Trader model has a sufficient expressive power.
The Trader has various binary operators as fundamental units, which allows it to represent any binary operations commonly used in practical trading strategies. Besides, the model also 
encompasses the linear models since we can choose the projection operator $(x, y) \mapsto x$ as $O_j$.

Ideally, we want to optimize the Traders by maximizing the cumulative returns:
\begin{eqnarray}\label{traderloss}
\Theta^* & \in \argmax_{\Theta} \sum_{u}\sign(f_\Theta(\vecret[1:S][0:u]))\cdot \ret[i][u+1] \nonumber \\
&:= \argmax_{\Theta} 
R(f_\Theta,\vecret[1:S][0:t],\ret[i][0:t+1])
\label{traderlossdef}
\end{eqnarray}
However, it is difficult to apply common optimization methods since the objective in the right-hand side is neither differentiable nor continuous w.r.t.~the parameter $\Theta$.
Therefore, we introduce a novel evolutionary algorithm driven by Company models, which we will describe below.

\subsection{Companies - Optimization and Aggregation Module}\label{sec:company}

As mentioned in the introduction, the behaviour of the financial market is highly unstable and uncertain, and thereby any single explanatory model is merely partially correct and transient. 
To overcome this issue and robustify the prediction, we develop a method to combine predictions of multiple Traders. This is in the same spirit as the general and long-standing framework of ensemble methods (e.g., \cite{hastie2009elements} or Chapter 4 of \cite{Timmermann2006}), but introducing an ``inductive bias'' that takes into account the dynamics of real-world financial markets would improve the performance of combined prediction. In particular, given the fact that the stock prices are determined as a result of diverse investments made by institutional traders, it is reasonable to consider a model imitating the environments in which institutional traders are involved (i.e., financial institutions).

\begin{algorithm}[ht]
    \caption{Prediction algorithm of Company}
    \label{companypredict}
    \algrenewcommand\algorithmicrequire{\textbf{Input:}}
    \algrenewcommand\algorithmicensure{\textbf{Output:}}
    \begin{algorithmic}[1]
    \Require $\vecret[1:S][0:t]$:stock returns before $t$,  Traders :$\{\Theta_n\}_{n=1}^N$
    \Ensure $\predret[i][t+1]$: predicted return of stock $i$ at $t$
    \Function{CompanyPrediction}{}
    \For{$n = 1,\cdots,N$}
        \State{$P_n \Leftarrow f_{\Theta_n}(\vecret[1:S][0:t])$
        }
        \Comment{Prediction by Trader \eqref{traders}}
    \EndFor\\
    \Return {$\mathrm{Aggregate}(P_1,\cdots ,P_N)$} \Comment{Aggregation}
    \EndFunction
    \end{algorithmic}
\end{algorithm}

\begin{algorithm}[ht]
    \caption{Educate algorithm of Company}
    \algrenewcommand\algorithmicrequire{\textbf{Input:}}
    \algrenewcommand\algorithmicensure{\textbf{Output:}}
    \label{companyeducate}
    \begin{algorithmic}[1]
    \Require $\vecret[1:S][0:t]$:stock returns before $t$
    \Require Traders. $N$ : the number of Traders. $Q$: ratio.
    \Ensure Traders
    \Function{CompanyEducate}{}
    \State{$R_n\Leftarrow R(f_{\Theta_n},\vecret[1:S][0:t],\ret[i][0:t+1])$}
    \Comment{Trader's return \eqref{traderlossdef}}

    \State{$R^* \Leftarrow$ bottom $Q$ percentile of $\{R_n \}$}
    \For{$n \in \{m| R_m \le R^* \}$}
        \Comment{for all bad traders}

        \State{Update $w_i$ in (\ref{traders}) by least squares method}
    \EndFor\\
    \Return{Traders}
    \EndFunction
    \end{algorithmic}
\end{algorithm}

\begin{algorithm}[ht]
    \caption{Prune-and-Generate algorithm of Company}
    \algrenewcommand\algorithmicrequire{\textbf{Input:}}
    \algrenewcommand\algorithmicensure{\textbf{Output:}}
    \label{companygen}
    \begin{algorithmic}[1]
    \Require $\vecret[1:S][0:t]$:stock returns before $t$, F: \# of fit times
    \Require $N$: the number of Predictors. $Q$: ratio.
    \Ensure $N'$ Predictors 
    \State{$\Theta_n \sim $ Unifrom Distribution}
    \For{$k=1,\cdots ,F$}
    \State{$R_n \Leftarrow R(f_\Theta,\vecret[1:S][0:t],\ret[i][0:t+1])$}
    \Comment{Trader's return \eqref{traderlossdef}}

    \State{$R^* \Leftarrow$ bottom $Q$-percentile of $\{ R_n \}$}
    \State{$\{\Theta_j\}_j \Leftarrow \{\Theta_n | R_n \ge R^* \}$}
    \Comment{Pruning}
    \State{$\{\Theta_j\}_{j=1}^{N'} \sim$ GM fitted to $\{\Theta_j\}_j$  *}
    \Comment{Generation}
    \EndFor\\
    \Return{$N'$ Predictors with $\{\Theta_j\}_{j=1}^{N'}$}
    \end{algorithmic}
    * If the parameter is an integer, we round it off.

\end{algorithm}

In our framework, a Company maintains $N$ Traders that act as weak learners or feature extractors, and aggregate them. Given $N$ Traders specified by parameters $\Theta_1, \ldots, \Theta_n$ and the past observations of stock returns $\vecret[1:S][0:t]$, a Company predicts the future returns by
\[
    \hat{r}[t + 1] =
    \mathrm{Aggregate}(f_{\Theta_1}, \ldots, f_{\Theta_n}).
\]
For clarity, this procedure is presented in Algorithm \ref{companypredict}. Here, $\mathrm{Aggregate}$ can be an arbitrary aggregation function and allowed to have extra parameters. For example, we can use the simple averaging $\frac{1}{N} \sum_{n = 1}^N f_{\Theta_n}(\vecret[1:S][0:t])$, linear regression or general trainable prediction models (e.g., neural networks and the Random Forest) that take the Traders' suggestions as the input features.

In order to achieve low training errors whilst avoiding overfitting, the Company should maintain the average quality as well as the diversity of the Traders' suggestions.
To this end, we introduce the Educate algorithm (Algorithm \ref{companyeducate}) and the Prune-and-Generate algorithm (Algorithm \ref{companygen}), which update weights and formulae of Traders, respectively.

\begin{itemize}
    \item \textbf{Educating Traders}: Recall that a single Trader \eqref{traders} is a linear combination of $M$ mathematical formula. A single Trader can perform poorly in terms of cumulative returns. However, if the Trader has good candidates of formulae (i.e., alpha factors), slightly updating the weights $\{ w_j \}$ while keeping the formulae would significantly improve the performance. Algorithm \ref{companyeducate} corrects the weights $\{ w_j \}$ of the Traders achieving relatively low cumulative returns. Here, we update the weights by the least-squares method, which is solved analytically.
    \item \textbf{Pruning poorly performing Traders and generating new candidate good Traders}: If a Trader holds ``bad'' candidates of formulae, keeping that Trader makes no improvement on the prediction performance while it can increase risk exposures. In that case, it may be beneficial to simply remove that Trader and replace it with a new promising candidate. Algorithm \ref{companygen} implements this idea. First, we evaluate the cumulative returns of the current set of Traders, and remove the Traders having relatively low returns. Then, we generate new Traders by randomly fluctuating the existing Traders with good performances.
    To this end, we fit some probability distribution to the current set of parameters, and draw new parameters from it.
    While the parameter specifying the formulae contain discrete variables such as indices of stocks and choices of arithmetic operations, we empirically found that fitting the Gaussian mixture distribution (i.e., a continuous multi-modal distribution) to discrete indices and discretizing the generated samples can achieve reasonably good performances. See Section \ref{sec:experiment} for detailed experimental results.
\end{itemize}

Using the above algorithms together, we can effectively search the complicated parameter space of Traders. Educate algorithm (Algorithm \ref{companyeducate}) is intended to be applied before pruning (Algorithm \ref{companygen}) to prevent potentially useful alpha factors from being pruned. In practice, given past observations of returns $\vecret[1:S][t_1: t_2]$, we can train the model by the following workflow.
\begin{enumerate}
    \item Educate a fixed proportion of poorly performing Traders by Algorithm \ref{companyeducate}.
    \item Replace a fixed proportion of poorly performing Traders with random new Traders by Algorithm \ref{companygen}.
    \item If the aggregation function $\mathrm{Aggregate}$ has trainable parameters, update them using the data $\vecret[1:S][t_1: t_2]$ and any optimization algorithm.
    \item Predict future returns by Algorithm \ref{companypredict}.
\end{enumerate}

We comment on some intuitions about the advantages of our method, although there is no theoretical guarantee in practical settings.
Algorithm \ref{companygen} increases the diversity of Traders by injecting random fluctuations to existing good Traders. From a generalization perspective, injecting randomness may help to avoid overfitting to the current transient state of the market. From an optimization perspective, we can view our algorithm as a variant of evolutionary algorithms such as the Covariance Matrix Adaptation Evolution Strategy (CMA-ES) \cite{Hansen1996}. While the original CMA-ES generates new particles from a Gaussian distribution, we see that using a multi-modal distribution is practically important. See Section \ref{sec:experiment} for an empirical verification.

\section{Experiments}\label{sec:experiment}

We conducted experiments to evaluate the performance of our proposed method on real market data. We compare our method with several benchmarks including simple linear models and state-of-the-art deep learning methods, and confirm the superiority of our method. We also demonstrate that our method can find profitable formulae (alpha factors) that are simple enough to interpret.

\subsection{Datasets}
Throughout the experiments, we used two real market datasets: (i) US stock prices listed on Standard \& Poor's 500 (S\&P 500) Stock Index and (ii) UK stock prices from the London Stock Exchange (LSE).
S\&P 500 and LSE have one of the highest trading volumes and market capitalization in the world.
From a reproducibility viewpoint, we used data that were distributed online free from Dukascopy's Online data feed\footnote{\url{https://www.dukascopy.com/}}.
For the S\&P 500 data, we used daily data for all the $500$ stocks listed S\&P 500 Stock Index in the period from May 19, 2000 to  May 19, 2020.
For the LSE data, we used hourly data for all the $77$ stocks prices available on Dukascopy in the period from September 07, 2016 to September 07, 2019.

\subsection{Evaluation Protocols}

\subsubsection{Time windows and execution lags}
In our experiments, we employed two practical constraints, time windows and execution lags.
In practice, we cannot use all the past observations of stock prices due to the time and space complexity. 
Also, we cannot trade stocks immediately after the observation of returns because we need some time to make an inference by the model and execute the trade.
Thus, it is reasonable to introduce a time window $w > 0$ and an execution lag $l > 0$, so we train models using observations $\vecret[i:S][t-l-w:t-l]$ and predict returns at time $t + 1$.
Throughout experiments, we used $w = 10$ and $l = 1$.

\subsubsection{Metrics}
To evaluate the performances of prediction algorithms, we adopted three metrics defined as follows. For each metric, higher value is better.
Let $r_i[t]$ be the return of stock $i$ ($i \in \{ 1, \ldots, S \}$) at time $t$, and let $\hat{r}_i[t]$ be its prediction obtained from an arbitrary method. Recall $\hat{b}_i[t] = \mathrm{sign}(\hat{r}_i[t)$.

\begin{itemize}
    \item \textbf{Accuracy (ACC)}
    the accuracy rate of prediction of the rise and drop of stock prices. $\text{ACC}_i=\mathrm{P}[\mathrm{sign}(\ret[i][t])=\predret[i][t]]$.
    \item \textbf{Annualized Return(AR)}: Given predictions of the returns of stock $i$, the cumulative return  is given as
    $C_i[t] := \sum_{u=0}^t \hat{b}_i[u + 1] r_i[u + 1]$.
    We define the Annualized Return \textbf{(AR)} averaged over all stocks as 
    $
    \mathrm{AR} := 100\times\frac{T_\mathrm{Y}}{T} \frac{1}{S} \sum_{i = 1}^S C_i[T]
    $
    , where $T_\mathrm{Y}$ is the average number of periods contained in one year.
    \item \textbf{Sharpe Ratio (SR)}: The Sharpe ratio \cite{sharpe1964capital}, or the Return/Risk ratio (R/R) is the return value adjusted by its standard deviation. That is, letting
    $\mu_i := \frac{1}{T} C_i[T]= \frac{1}{T} \sum_{t = 1}^T \hat{b}_i[u + 1] r_i[u + 1]\quad \text{and}
    \sigma_i^2 := \frac{1}{T} \sum_{t = 1}^T (\hat{b}_i[u + 1] r_i[u + 1] - \mu_i)^2,$
    and we define $\mathrm{SR}_i := \mu_i / \sigma_i$. Then, we report the average $\mathrm{SR} = \frac{1}{S} \sum_{i=1}^S \mathrm{SR}_i$.
    \item \textbf{Calmar Ratio (CR)}: We also use the Calmar ratio \cite{young1991calmar}, another definition of adjuster returns. Define the Maximum DrawDown (MDD) as
    \[
        \textstyle
        \mathrm{MDD}_i := \max_{1 \leq t \leq T}
        \max_{t < s \leq T} \left(
            1 - \frac{C_i[t]}{C_i[s]}
        \right).
    \]
    The Calmar ratio is defined as $\mathrm{CR}_i := \mathrm{AR}_i / \mathrm{MDD}_i$. In our experiments, we report $\mathrm{CR} = \frac{1}{S} \sum_{i=1}^S \mathrm{CR}_i$. Note that while both SR and CR are adjusted returns by its risk measures, CR is more sensitive to drawdown events that occur less frequently (e.g., financial crises).
\end{itemize}

\subsection{Baseline Methods}\label{sec:baseline}
We compared our methods with the following baseline methods:
\begin{itemize}
    \item Market: a uniform Buy-And-Hold strategy.
    \item Vector Autoregression (VAR): a commonly-used linear model for financial time series forecasting \cite{Sims1980}.
    \item Random Forest (RF): a commonly-used ensemble method \cite{Breiman2001}.
    \item Multi Head Attention (MHA): a deep learning algorithm for time series prediction \cite{Vaswani2017}.
    \item Long- and Short-Term Networks (LSTNet): a deep-learning-based algorithm which combines Convolutional and Recurrent Neural Network \cite{Lai2018}.
    \item State-Frequency Memory Recurrent Neural Networks (SFM)\footnote{There was an unintended data leak in the implementation published by the author. Therefore, we fixed it in our experiment for fairness.}: a deep learning-based stock price prediction algorithm that incorporates the concept of Fourier Transform into Long Short-Term Memory
    \cite{Zhang2017}.
    \item Symbolic Regression by Genetic Programming (GP): a predicton algorithm using genetic programming \cite{Poli2008}.
\end{itemize}

To verify the effects of individual technical components in our proposed method (TC), we also compared the following ``ablation'' models.
\begin{itemize}
    \item Changing the Trader structure: \textbf{TC linear} only uses the linear activation function $x \mapsto x$, so the eventual prediction of a Campany becomes just a linear combination of several binary operations. \textbf{TC unary} only uses the unary operator $O(x, y) = x$.
    \item Changing the optimization algorithm: \textbf{TC w/o educate} does not execute the Educate Algorithm (Algorithm \ref{companyeducate}), so Traders can be discarded even if they have promising formulae. \textbf{TC w/o prune} does not execute the pruning step in Algorithm \ref{companygen}, so a Company keeps poorly performing Traders. \textbf{TC unimodal} uses the Gaussian distributions instead of the Gaussian mixtures in the generation step. \textbf{TC MSE} uses the mean squared loss as scores for educating and pruning, so it does not use the cumulative returns at all.
\end{itemize}

Table \ref{tab:hyperparamters} lists the hyper-parameters used in the baseline algorithms.

\begin{table}[tb]
\caption{Hyper-parameters in Experiments}
\label{tab:hyperparamters}
{\small
\begin{tabular}{ccc}
    \hline
    \textbf{Parameter} & \multicolumn{1}{c}{\textbf{Value}} & \textbf{Definition} \\
    \midrule
    $\numterm$ &  $\{1,\cdots,10\}$  & \eqref{traders}\\
    $D_j,F_j$ & $\{0,\cdots,10\}$ & \eqref{traders}\\
    $A_j(x) $ &  $\{x,\mathrm{tanh}(x),\mathrm{exp}(x),\mathrm{sign}(x),\mathrm{ReLU}(x)\}$ & \eqref{traders}\\
    $O_j(x,y)$ & \begin{tabular}{@{}c@{}}$\{x+y, x-y , xy, x, y,\max(x,y)$, \\ $ \min(x,y),x>y,x<y,\textrm{Corr}(x,y) \}$\end{tabular}  & \eqref{traders}\\
    $N$ & $100$ & Algorithm \ref{companypredict} \\
    Aggregate & Linear regression & Algorithm \ref{companypredict} \\
    $Q$ & $0.5$ & Algorithm \ref{companygen} \\
    \hline
\end{tabular}
}
\end{table}

\subsection{Performance Evaluation on Real Market Data}

\subsubsection{Offline prediction}
First, we trained the models using the first half of the datasets, and then evaluated the performances of the models with frozen parameters using the latter half. For US market, we used the data before May 2018 for training and the rest for testing. For UK market, we used the first one and a half years for training and the rest for testing.

\begin{figure}[tb]
    \centering
    \includegraphics[width=0.9\columnwidth]{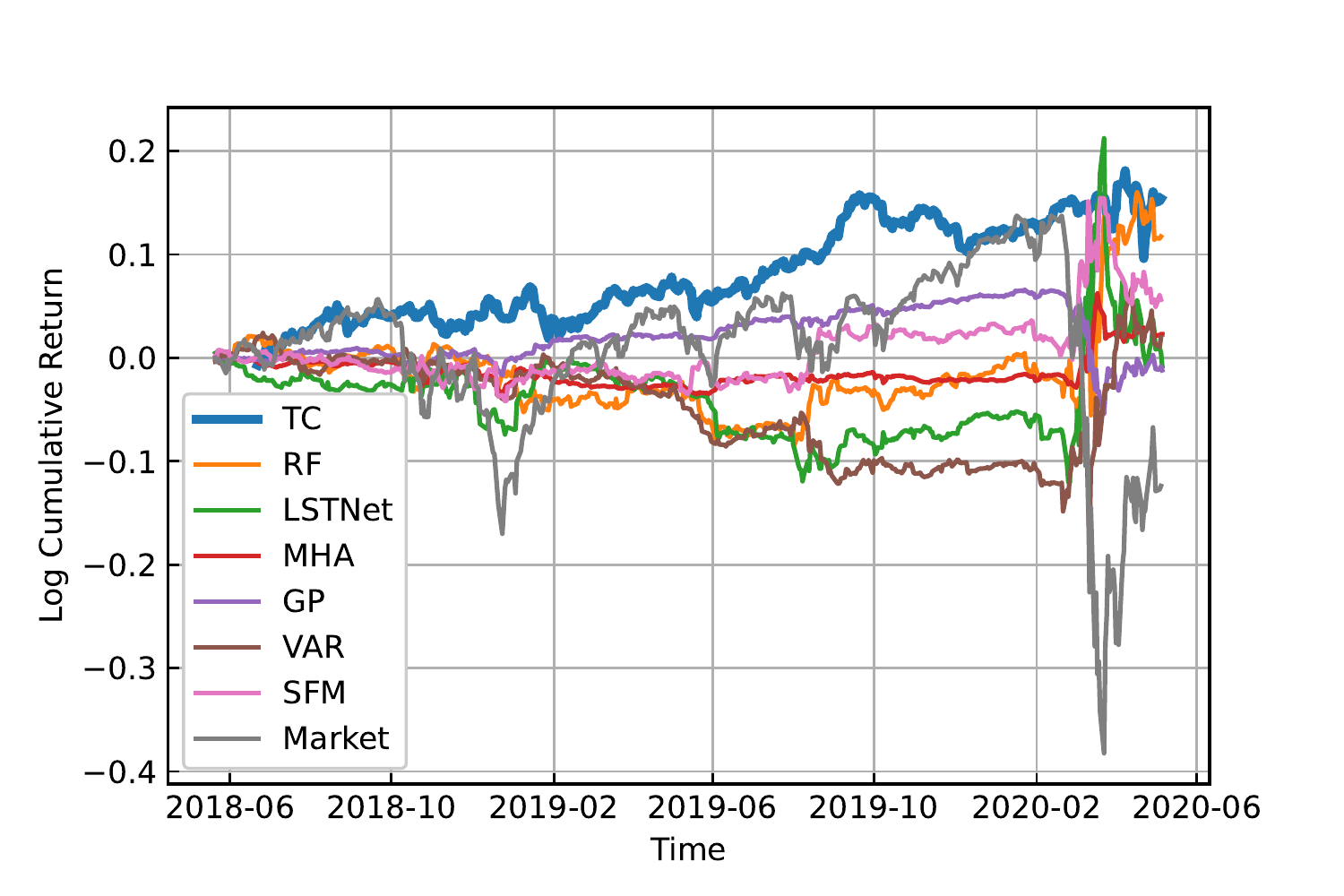}
    \caption{Cumulative returns on US market}
    \label{fig:offline_us}
    \Description{A figure showing the cumulative returns achieved by several methods on US market. The horizontal and vertical axes are the time and the log-cumulative return, respectively.
    }
\end{figure}

\begin{figure}[tb]
    \centering
    \includegraphics[width=0.8\columnwidth]{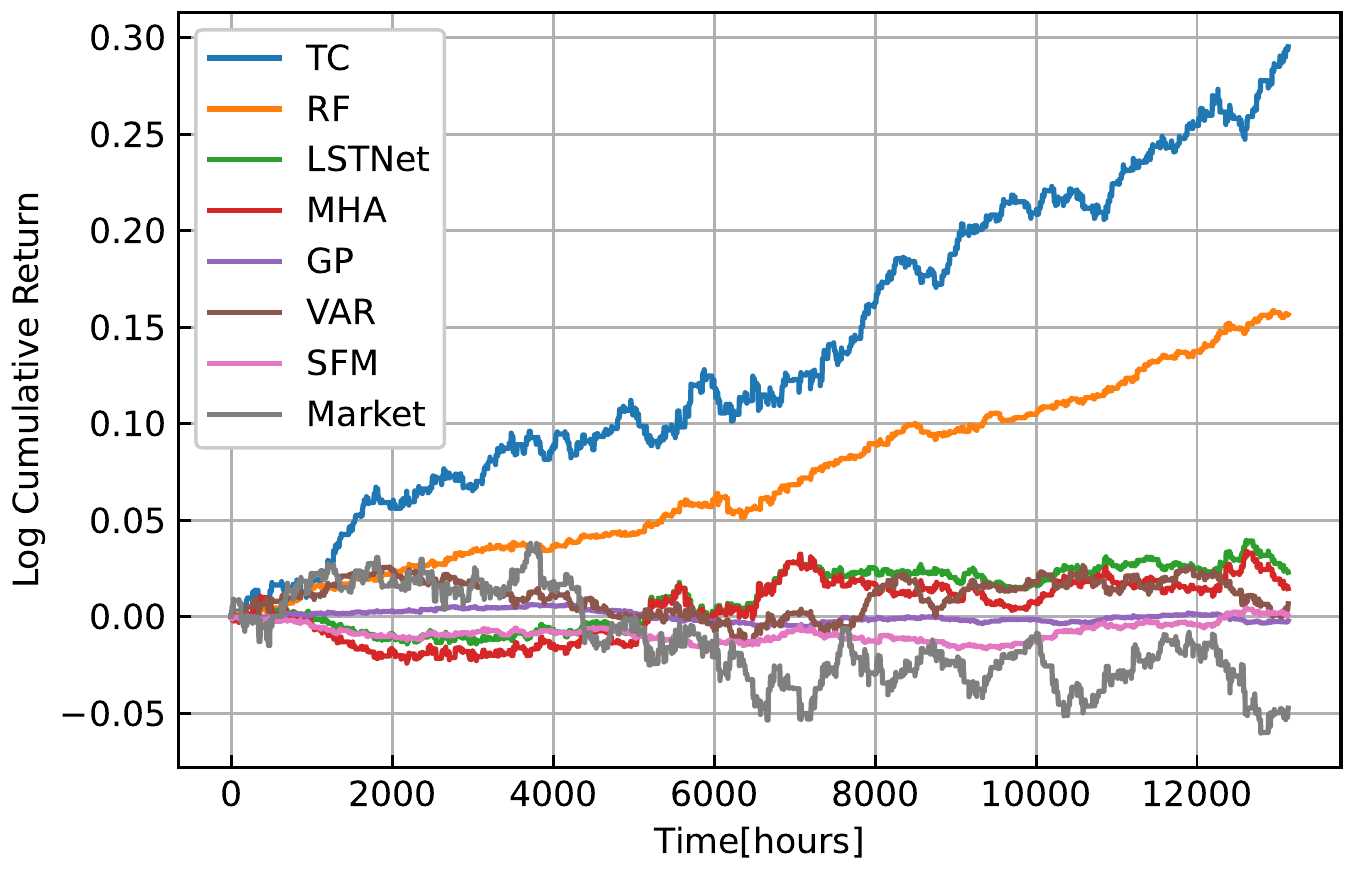}
    \caption{Cumulative returns on UK market}
    \label{fig:offline_uk}
    \Description{A figure showing the cumulative returns achieved by several methods on UK market. The horizontal and vertical axes are the time and the log-cumulative return, respectively.
    }
\end{figure}

\begin{table}[tb]
    \caption{Performance comparison on US markets}
    \label{tab:us}
    \centering
    {\small
    \begin{tabular}{lcccc}
\hline
                    &  ACC(\%)  &         AR(\%) &            SR &             CR \\
\midrule
            {\bf Market} & 55.09  & -5.45  &   -0.2 &  -0.12  \\
               {\bf VAR} & 49.86  & 3.51  &  0.28  &   0.24 \\
               {\bf MHA} & 48.16 $\pm$ 1.1 & 0.66 $\pm$ 0.54 &   0.12 $\pm$ 0.10 &   0.11 $\pm$ 0.09 \\
            {\bf LSTNet} & 47.51 $\pm$ 1.12 & 3.61 $\pm$ 2.09 &   0.18 $\pm$ 0.10 &    0.16 $\pm$ 0.1 \\
               {\bf SFM} & 50.14 $\pm$ 1.09 &  5.95 $\pm$ 1.16 &   0.52 $\pm$ 0.10 &    0.50 $\pm$ 0.18 \\
                {\bf GP} & 54.99 $\pm$ 1.15  & -0.27 $\pm$ 0.67 &  -0.03 $\pm$ 0.10 &  -0.02 $\pm$ 0.08 \\
                {\bf RF} &  53.74 $\pm$ 1.11 & 4.38 $\pm$ 1.60 &  0.28 $\pm$ 0.11 &   0.25 $\pm$ 0.13 \\ \hline
 {\bf TC linear} & 53.23 $\pm$ 1.18  & 2.81 $\pm$ 0.36 &   0.86 $\pm$ 0.10 &   0.93 $\pm$ 0.28 \\
   {\bf TC unary} & 51.14 $\pm$ 1.17 &  1.50 $\pm$ 0.33 &  0.46 $\pm$ 0.11 &   0.44 $\pm$ 0.17 \\
       {\bf TC w/o educate} & 52.36 $\pm$ 1.15 &  3.69 $\pm$ 0.32 &   1.12 $\pm$ 0.10 &   1.36 $\pm$ 0.38 \\
      {\bf TC w/o prune} &   50.31 $\pm$ 1.14  & 1.08 $\pm$ 0.33 &  0.34 $\pm$ 0.11 &    0.30 $\pm$ 0.13 \\
    {\bf TC unimodal} & 50.33 $\pm$ 1.17 &  0.08 $\pm$ 0.35 &  0.02 $\pm$ 0.11 &   0.02 $\pm$ 0.08 \\
  {\bf TC MSE} & 51.35 $\pm$ 1.19 & 1.56 $\pm$ 0.54 &   0.30 $\pm$ 0.11 &   0.26 $\pm$ 0.13 \\
                {\bf TC (Proposed)} & {\bf 55.68 $\pm$ 1.16} &  {\bf 10.72 $\pm$ 0.86} &  {\bf 1.32 $\pm$ 0.11} &    {\bf 1.57 $\pm$ 0.44} \\
\hline
\end{tabular}
    }
\end{table}

\begin{table}[tb]
    \caption{Performance comparison on UK markets}
    \label{tab:uk}
    \centering
    {\small
    \begin{tabular}{lcccc}
    \hline 
    & {\bf ACC(\%)} & {\bf AR(\%)} &{\bf R/R}&{\bf CR}  \\
    \midrule
    {\bf Market} & 50.009 & -3.34 & -0.177 & -0.076  \\
    {\bf VAR} & 49.965 & 0.54 & 0.064 & 0.031  \\
    \hline
    {\bf MHA} & 49.943 $\pm$ 0.991  & 1.01 $\pm$ 2.91 & 0.04 $\pm$ 0.15 & 0.03 $\pm$ 0.05  \\
    {\bf LSTNet}& 49.997 $\pm$ 1.007   & 1.53  $\pm$ 3.29  & 0.07 $\pm$ 0.18 & 0.05 $\pm$ 0.08  \\
    {\bf SFM} & 50.472 $\pm$ 8.810  & 0.23 $\pm$ 4.23 & 0.00 $\pm$ 0.22 & 0.02 $\pm$ 0.10  \\
    {\bf GP} & 50.017 $\pm$ 2.128  & -0.09 $\pm$ 1.47 & -0.02 $\pm$ 0.36 & 0.05 $\pm$ 0.12  \\
    {\bf RF} & 50.719 $\pm$ 1.177 & 10.45 $\pm$ 1.98 & 1.23 $\pm$ 0.23 & 1.17 $\pm$ 0.56  \\
    {\bf TC} & {\bf 50.928 $\pm$ 1.115 } & {\bf 32.32 $\pm$ 1.04 } & {\bf 2.23 $\pm$ 0.07} & {\bf 3.32 $\pm$ 0.44} \\
  \hline
\end{tabular}
    }
\end{table}

Table \ref{tab:us} and Table \ref{tab:uk} show the comparisons of between our proposed method (TC) and other baseline methods on US and UK markets, respectively. All methods are evaluated using three evaluation metrics (AR, SR, and CR). For methods depending on random initializations, we run the evaluations for each method 100 times with different random seeds and provide the means and the standard deviations. Also, Figure \ref{fig:offline_us} and Figure \ref{fig:offline_uk} show the cumulative returns on US and UK markets, respectively. 

Overall, our method outperformed the other baselines in the presented three evaluation metrics. Some interesting observations are as follows.

\noindent \textbf{Importance of Traders}: \
Comparing several baseline methods and ablation models, we found that the structure of Traders is of crucial importance.

First, in the definition of Traders \eqref{traders}, we restrict the formulae to those represented by the binary operators $O_j$. This means that Traders rule out formulae that leverage interactions between three or more terms, which can reduce the complexity of the entire model without losing the expressive power. This is corroborated by the following observations: Among the baseline methods, a simple linear method (VAR) achieved a relatively good performance.
VAR estimates its coefficients by the ordinary least-squares method, which means that the prediction of the return of each individual stock can be a ``dense'' linear combination of past observations. On the other hand, TC linear, which improved SR significantly upon VAR, finds the solution among linear combinations of features made of at most two observations. On the other hand, we can also see that using only unary operations (TC unary) greatly deteriorates the performance.

Second, comparing TC and TC linear, we see that introducing non-linear activation functions also improves the performance.

Third, TC also outperformed another off-the-shelf ensemble method (RF). RF combines non-linear predictors given by decision trees, i.e., indicator functions of rectangles (see e.g., Chapter 9 of \cite{hastie2009elements}). RF requires many decision trees to approximate binary operations such as $\max{x, y}$ or $x > y$. Hence, when these operations are actually important for constructing alpha factors, RF can increase the redundancy and the model complexity, which leads to poor performance especially in SR and CR. In fact, these operations frequently appear in good Traders (Table \ref{tab:formula}).

\begin{table}[t]
\centering
\caption{Prediction formula extracted from the best Traders.
These expressions are for predicting the returns at $t+1$
}
\label{tab:formula}
\begin{tabular}{cl}
    \hline
    \# of terms  & ${\rm LLOY}_{t+2}$ \\
    \midrule
    1 & \begin{tabular}{l} $=-2.28({\rm SHP}_{t}>{\rm AHT}_{t-3})$ \end{tabular}\\
    \hline
    2 & \begin{tabular}{l}
    $=-0.80\sign({\rm WPP}_{t-3}+{\rm SKY}_t)$\\
    $-1.85\sign(\min({\rm AV}_{t-2},{\rm SHP}_{t-5}))$
    \end{tabular}\\
    \hline
    3 & \begin{tabular}{l}
    $=4.919{\rm ReLU}({\rm AHT}_{t-3})$\\
    $+5.859\sign(\max({\rm WEIR}_{t-2},{\rm WTB}_{t-5}))$\\
    $-1.07({\rm PFC}_{t-1}>{\rm DGE}_{t-3})$ \\
    \end{tabular}
    \\
  \hline
 \end{tabular}
\end{table}

\noindent \textbf{Importance of combining multiple formulae}: \
Among the baseline methods, GP outputs a single mathematical formula by using genetic programming \cite{Poli2008}. However, this did not work well in our experiments in which we adopted reasonably long test periods. This may reflect the fact that any single formula is ephemeral due to the (near) efficiency of the markets. On the other hand, our method that maintains multiple formulae performed well over the test period.

\noindent \textbf{Importance of optimization heuristics}: \
We found that each individual optimization technique presented in Section \ref{sec:company} significantly improves the performance.
First, the pruning step seems quite important (cf. TC w/o prune). Regarding the scores for the pruning, using the MSE instead of the cumulative returns deteriorates the performance (cf. TC MSE). Second, we can see that introducing the education step also improves the overall performance (cf. TC w/o educate). Otherwise Companies may discard Traders that have possibly good formulae. Lastly, using multimodal distribution in the generation step (Algorithm \ref{companygen}) is quite important. If we instead use a unimodal distribution (cf. TC unimodal), the performance is substantially deteriorated. A possible reason is that a unimodal distribution concentrates around the means of discrete indices, which does not make sense.

\noindent \textbf{Comparison to other non-linear predictors}: \
We also compared our method to other complex non-linear models. MHA, LSTNet and SFM are prediction methods based on deep learning. Among these, SFM performed relatively well in US market, but none of them achieved good performances in UK market. Our method consistently outperformed these methods.

\subsubsection{Online prediction}

\begin{figure}[tb]
    \centering
    \includegraphics[width=0.9\columnwidth]{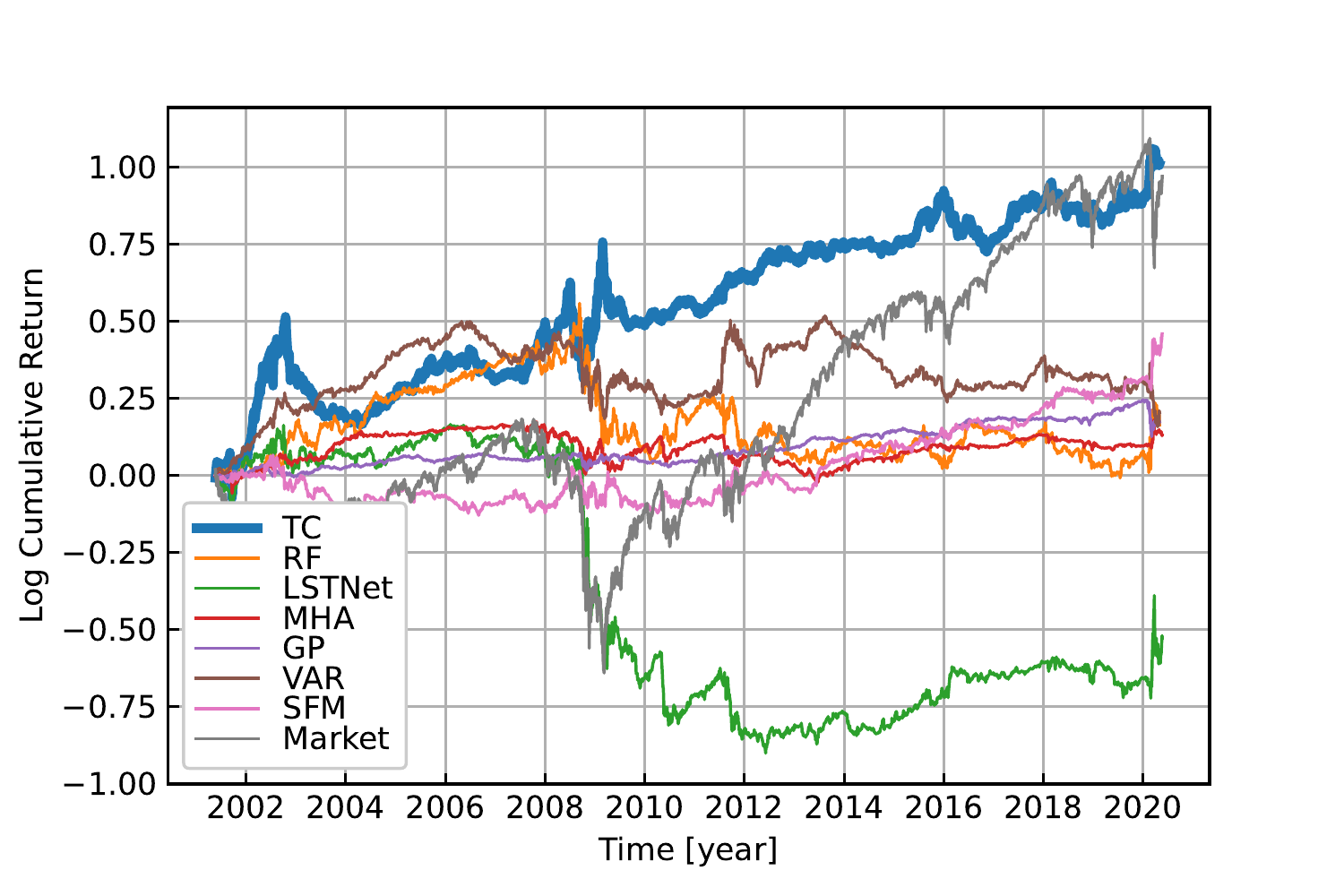}
    \caption{
        Cumulative returns on US market in the online prediction setting
    }
    \label{fig:online_us}
    \Description{A figure showing the cumulative returns achieved by several methods on US market. The horizontal and vertical axes are the time and the log-cumulative return, respectively.
    }
\end{figure}

In the previous experiment, we adopted a simple train/test splitting. However, in practice, it is not reasonable to use a single model for a long time, and we might update the model more frequently to follow structural changes of the markets. Here, using the US data, we also evaluated our method in a sequential prediction setting. We sequentially updated the models every year, where we used all the past observations for training. Figure \ref{fig:online_us} shows the cumulative returns on US market. Our method performed well also in this setting.

\subsection{On Interpretablity of Traders}\label{sec:disc_interp}

\begin{figure}[t]
    \centering
    \includegraphics[width=0.8\columnwidth]{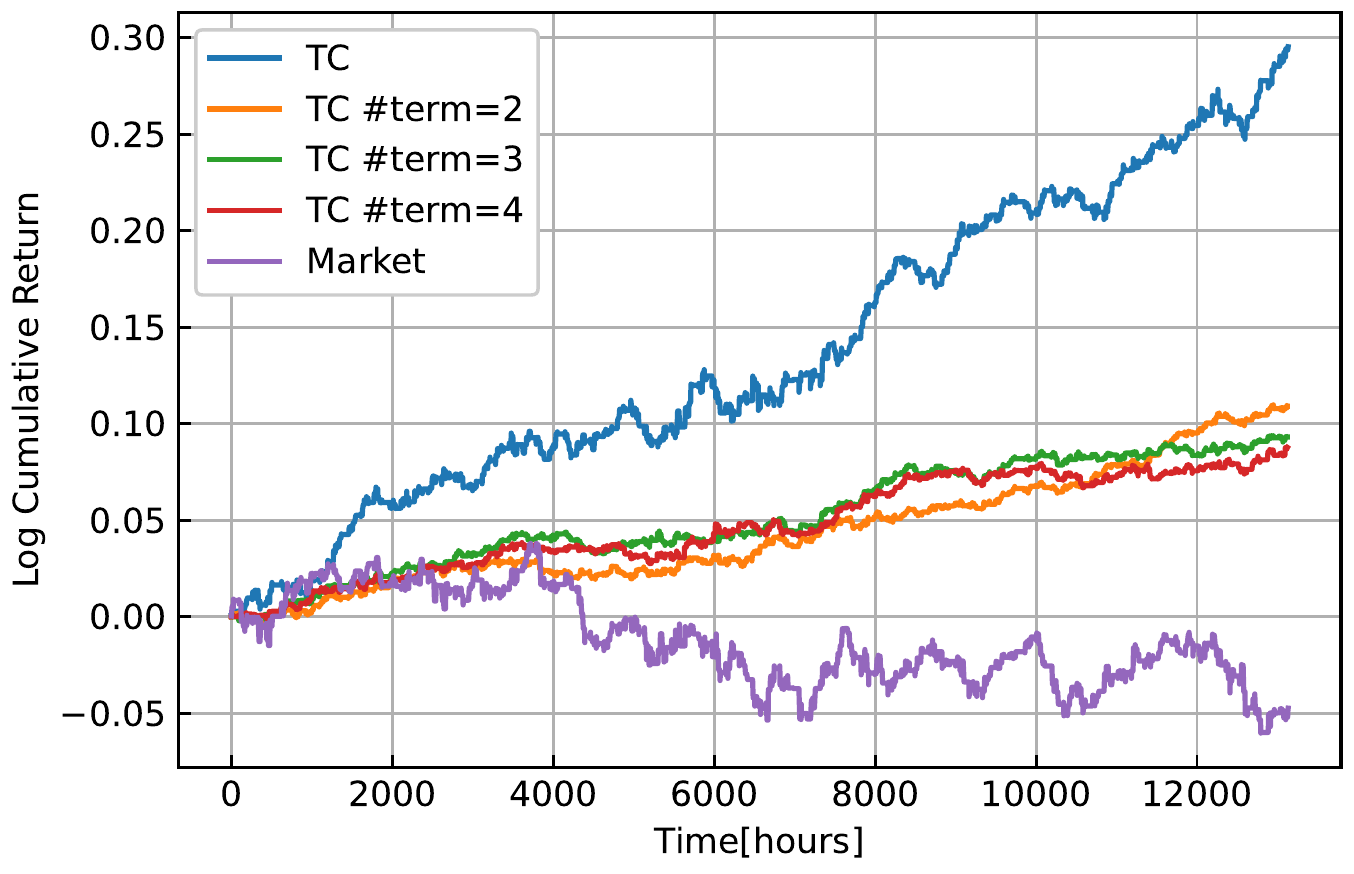}
    \caption{Cumulative returns achieved by individual Traders extracted from Companies. ``TC'' is the overall performance of our method.
    ``TC \# term = M'' stands for a single best performing Trader extracted from a Company with fixed parameter $M$.}
    \label{fig_interp}
    \Description{A figure showing the cumulative returns achieved by several methods on UK market. The horizontal and vertical axes are the time and the log-cumulative return, respectively.
    }
\end{figure}

\begin{figure}[t]
\centering
\includegraphics[width=0.8\columnwidth]{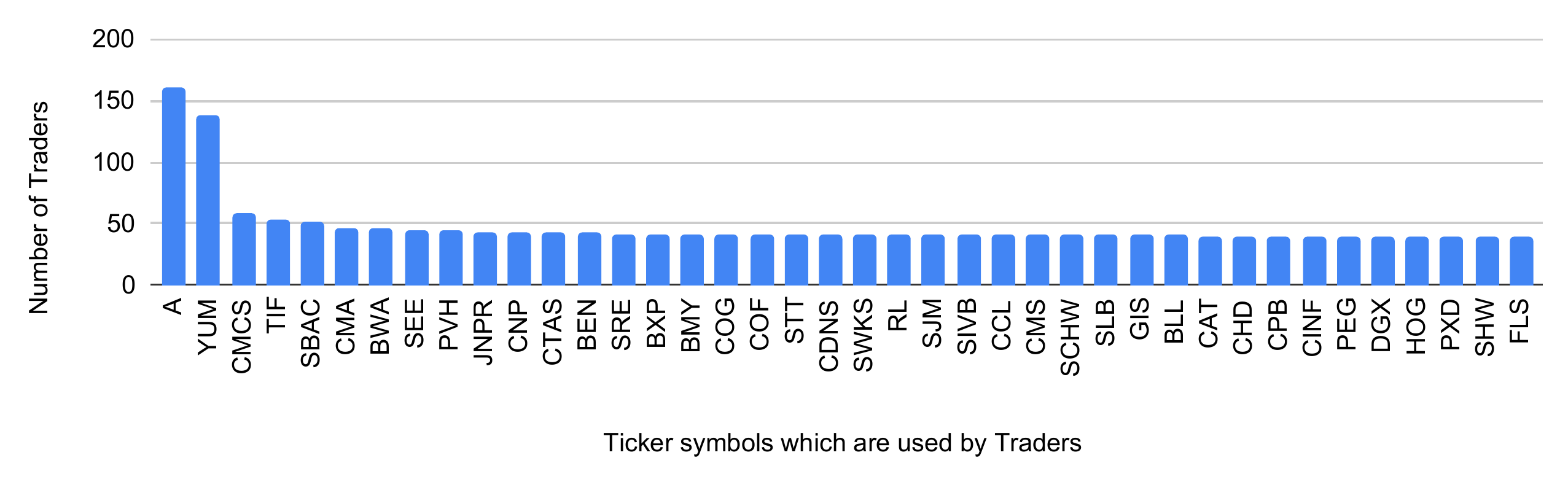}\\
\includegraphics[width=0.8\columnwidth]{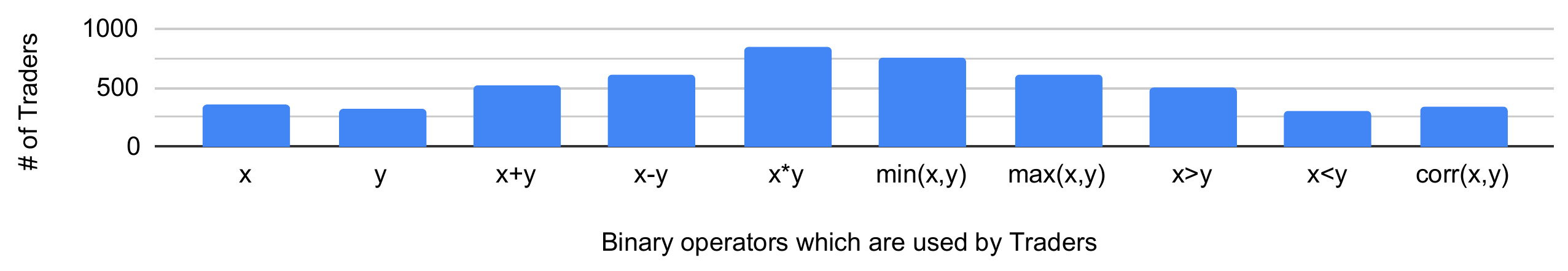}
\caption{The number of times each parameter was used.(Upper) indices $P_j, Q_j$. (Lower) operators $O_j$}
\label{fig_trader_count}

\end{figure}

So far, we have evaluated the overall performance of the proposed method. Here, we investigate the performance of a single Trader or a formula extracted from the trained model, and discuss about the interpretation of obtained formulae.

First, we investigate the performance of a single \textit{Trader}. Recall that a Trader \eqref{traders} is given as a linear combination of $M$ mathematical formulae. Thus, the expressive power of a single Trader increases as $M$ increases. Here, using the UK data, we trained the Companies by restricting $M$ to be a fixed value in $\{ 1, 2, 3, 4 \}$ (Note that, in our method, $M$ for each Trader is trainable by default). Then, we extracted the best performing Trader from each trained Company, and evaluated the performance for the test period. Figure \ref{fig_interp} shows the result. We can see that our method can find Traders that achieve positive returns by themselves, while the total return of the overall market is negative during the test period. If we increase the number of terms $M$, the cumulative return of a single Trader at the end of the test period decreased (albeit slightly), which suggests that increasing the expressive powers of individual Traders is prone to overfitted formulae and does not necessarily result in a single profitable formula. Meanwhile, we should note that the overall performance of the Company (TC in Figure \ref{fig_interp}) is much better than a single Trader.

Next, we consider the interpretability issue.
Table \ref{tab:formula} lists the actual formulae extracted from the trained Traders.
Regarding the meanings of the ticker symbols used in the formulae, see the website of London Stock Exchange \footnote{\url{https://www.londonstockexchange.com/}}.
For example, ${\rm{LLOY}_{t+2}}$ indicates the return from $t+1$ to $t+2$ of Lloyds Bank.
We believe that each formula would be informative for real-world investors and can provide useful insights into their investments.
Notably, we found that the extracted formulae often contain binary operations such as $\min(x, y)$, $\max(x, y)$ and pairwise comparisons ($x > y$). These operators are difficult to be approximated by decision trees that can only represent rectangular regions, which may be the reason for the superiority of our method over the Random Forest.

Figure \ref{fig_trader_count} shows which ticker symbol and binary operator are used by Traders and how many times to predict the return of AAPL (Apple Inc.). 
From Figure \ref{fig_trader_count}, we can interpret which stocks the Company is paying attention to and binary operators the Traders are using.

\section{Related work}

\subsection{Financial Modelling and Machine Learning}

There are two established approaches to financial time series modeling. Firstly, many \textit{statistical time series models} that aim to describe the generative processes of financial time series have been developed. For example, the autoregressive integrated moving average (ARIMA) model and Vector autoregression (VAR) are often used in financial time series prediction \cite{Box2015,Ltkepohl2005}.
Secondly, another direction is to use \textit{factor models}, which aim to explain asset prices using not only the structure of each individual time series but also the cross-section information and other kinds of financial information. To name a few, Fama--French's three-factor model \cite{fama1992}, Carhart four-factor model \cite{Carhart1997}, and Fama-French's five-factor model \cite{fama2015} are among the most important ones. More than 300 identified factors are reported as of 2012 \cite{Harvey2015}.

Recent financial econometrics is characterized by the combination of financial modeling and machine learning methods, especially deep learning.
Although there are many research directions in this flourishing field, these include directions that (i) apply sophisticated time series models to financial problems and (ii) incorporate data of different modalities into the prediction.
Sezer et al.~conducted an extensive survey of these applications \cite{Sezer2019}. 
For the first direction, Zhang et al.~\cite{Zhang2017} proposed to used frequency information to forecast stock prices.
Lai et al.~\cite{Lai2018} proposed a method to extract long-term and short-term patterns by combining CNN and RNN.
For the second direction, the use of news, social media and networks among companies is also active \cite{Hu2018,Xu2018,Chen2018,Ding2015}.

However, it is folklore among experts that, under the transient and uncertain environments of financial markets, complex models (including neural networks) do not work well as expected, and traditional simpler models are more preferred. In partcular, Makridakis et al.~\cite{Makridakis2018} pointed out the superiority of ``traditional'' statistical models over machine learning models in financial time series. This motivates us to leverage simple units of models such as formulaic alphas \cite{Kakushadze2018, tulchinsky2015alphas} and deal with the uncertainty by bootstrapping them, instead of using ``black-box'' deep learning-based methods.

\subsection{Ensemble Methods}

Combining multiple predictions is a long-standing approach in data science \cite{Timmermann2006, hastie2009elements}. Generally speaking, model selection and model ensemble are both important ideas (e.g., Chapter 8 of \cite{hastie2009elements}). However, in some situations, selecting only a single model with the (temporal) best track record can lead to suboptimal performances, which has been confirmed empirically (e.g., Section 7.2 of \cite{Timmermann2006}) and theoretically (e.g., \cite{juditsky2008}). Therefore, in many situations, ensemble-type methods might be the first candidate to try.

In ensemble methods, some techniques such as pruning and random generation of experts have shown to be effective in various situations. For example, it has been widely known that eliminating poorly performing experts (e.g., \cite{friedman1991} and Section 7.3 of \cite{Timmermann2006}) or partial structures of experts (e.g., \cite{breiman1984}) can improve the overall performance. The idea of random generation of experts has been used in the Random Forest or the Random Fourier Features method \cite{rahimi2008rf}. 
We would like to stress that, as we demonstrated in Section \ref{sec:experiment}, the designs of pruning and generation schemes are crucially important in financial time series prediction.

\subsection{Metaheuristics in Finance}
Over the years, many research have been done on the application of metaheuristics to finance \cite{Allen1999}.
Soler-Dominguez et al.~\cite{SolerDominguez2017} has done an extensive survey on these application.
While portfolio optimization and index tracking and enhanced indexation are active applications of metaheuristics, the application of Genetic Programming (GP) is common for stock price prediction.
Index prediction method, combination with self-organizing map (SOM), one using multi-gene and one using hybrid GP were proposed \cite{Sheta2015,Manahov2015,Hsu2011}.

\section{Conclusion}
We proposed a new prediction method for financial time series.
Our method consists of two main ingredients, the Traders and the Company. The Traders aim to predict the future returns of stocks by simple mathematical formulae, which can be naturally interpretable as ``alpha factors'' in finance literature. The Company aggregates the predictions of Traders to overcome the highly uncertain environments of financial markets. The Company also provides a novel training algorithm inspired by real-world financial institutes, which allows us to search over the complicated parameter space of Traders and find promising mathematical formulae efficiently.
We demonstrated the efficacy of our method through experiments on US and UK market data. In particular, our method outperformed some common baseline methods in both markets, and an ablation study showed that each individual technique in our proposed method does improve the overall performance.
We focused on forecasting stock prices throughout this paper, and an interesting future direction is to investigate the applicability of our method to other types of assets.

\begin{acks}
We thank the anonymous reviewers for their constructive suggestions and comments. We also thank Masaya Abe, Shuhei Noma, Prabhat Nagarajan and Takuya Shimada for helpful discussions.
\end{acks}

\bibliographystyle{ACM-Reference-Format} 
\bibliography{main}

\end{document}